\documentclass[11pt]{article}
\usepackage{moriond,epsfig}

\def\be{\begin{equation}}
\def\ee{\end{equation}}
\def\bea{\begin{eqnarray}}
\def\eea{\end{eqnarray}}

\begin{document}
\vspace*{4cm}
\title{D\O\ STATUS AND FIRST RESULTS FROM RUN 2}

\author{AURELIO JUSTE \\ (on behalf of the D\O\ Collaboration)}

\address{Fermi National Accelerator Laboratory, P.O. Box 500,
Batavia, IL 60510, USA}

\maketitle\abstracts{
In order to fully exploit the physics potential of the
Tevatron Run 2, the D\O\ detector has been upgraded. Having
nearly completed the commissioning phase, the D\O\ detector is
starting to produce its first physics results. An overview of the
status of the main subdetectors involved in the upgrade is given, followed
by some examples of preliminary physics results already emerging.}
\section{Introduction}
Run 2 of the upgraded Tevatron collider started in March 2001. The
ultimate goal is to accumulate $\sim$2 fb$^{-1}$ of integrated luminosity
by 2005, which represents a
twenty-fold increase with respect to Run 1. 
The center-of-mass energy has been increased from 1.8 TeV to 1.96
TeV and the bunch crossing interval reduced from 3.5 $\mu$s to 396
ns. In order to fully exploit the physics opportunities of the
high luminosity Main Injector era, as well as to be able to cope
with the higher event rates and backgrounds expected, the D\O\
detector has been upgraded \cite{d0upgr}. Run 2 offers an
extremely exciting physics program which includes high statistics
studies on the top quark sector, searches for the Higgs boson and
new phenomena,
precision electroweak measurements, b-physics and QCD.

\section{Status of the D\O\ Detector}
The upgrade builds on the strengths of the D\O\ detector, namely,
excellent calorimetry and muon system and, most importantly, augments its tracking
and triggering capabilities: new inner tracking system, new
central and forward preshower detectors and a new pipelined 3-level trigger system. The
inner tracking consists of a scintillating fiber tracker and a
silicon microstrip tracker, embedded in a 2 Tesla axial magnetic
field, which is provided by a $\sim$2.6 m long super-conducting
solenoid with $\sim$0.5 m inner radius.

The D\O\ detector was rolled into the collision hall in January
2001, with the first collisions being delivered in April 2001.
During the summer 2001, the emphasis was put in commissioning and
timing in the detector, as well as improving the electronics, data
acquisition and offline reconstruction. In October 2001, during a
six week shutdown, most of the central fiber tracker electronics
were installed. As of March 2002, the accelerator has delivered
$\sim 27$ pb$^{-1}$ of integrated luminosity, of which $\sim 10$
pb$^{-1}$ have been collected by D\O. A significant fraction
($\sim25\%$) of this integrated luminosity has been devoted to
detector commissioning.

An overview of the status of the main subdetectors involved in
the upgrade is given below.
\subsection{Silicon Microstrip Tracker (SMT)}
The SMT is the innermost tracking element with a total of
$\sim800,000$ readout channels. It consists of six barrel modules
with silicon sensors parallel to the beamline, as well as a total of 16
disks (four of them interspersed with the barrels) with silicon sensors normal to the
beamline for forward tracking ($\mid\eta\mid\leq 3$). 
The barrels are 12.4 cm long and contain four concentric
layers with radii ranging from 2.7 cm to 9.4 cm. The use of single- and
double-sided (2$^{\rm o}$ or 90$^{\rm o}$ stereo view) sensors
provides 3-D track reconstruction capabilities. The
detector is fully commissioned, with the barrels and 12 innermost disks
(4 outermost disks) being $\sim95\%$ ($\sim86\%$) operational.
\subsection{Central Fiber Tracker (CFT)}
The CFT spans the radial region $\rm 20<r<51$ cm and consists of
eight concentric barrels of axial and stereo scintillating fiber
doublets. The fibers, with 830 $\mu$m diameter, are read out through
$\sim12$ m long clear wave-guides down to Visible Light Photon
Counters (VLPCs), a variant of the solid-state
photomultiplier. VLPCs operate at 9K, and
have high quantum efficiency ($\sim85\%$) and excellent
signal-to-noise ratio. The total number of readout channels is
77,000. The CFT also provides a fast track trigger at Level 1. The
subdetector is close to being fully commissioned, with only missing
readout boards for 48$\%$ of the stereo view. It is expected to be
fully instrumented by mid-April 2002.
\subsection{Calorimeter}
The upgrade preserves the excellent Run 1 calorimeter, based on LAr
sampling with U absorber, which is rather uniform in response, 
nearly compensating and has a good energy resolution and coverage
up to $\mid\eta\mid\leq 4.2$. The front-end electronics is upgraded in order to
accommodate for the reduced bunch spacing, while preserving the noise
performance. The subdetector is fully commissioned and performing very well, 
with only $\sim 0.1\%$ of a total of 55,000 readout channels being dead or noisy.
\subsection{Muon System}
The upgrade is driven by the goal of maximizing the acceptance for
muons from high-p$_T$ processes, which implies extending the detector coverage up to 
$\mid\eta\mid\leq 2$ and having efficient, unprescaled triggers. Like for the
calorimeter, the front-end electronics has also been upgraded. Tracking is performed
by three layers of wire chambers, whose
hits must be verified by corresponding coincidences in three layers of trigger
scintillators with good granularity ($\Delta\eta\times\Delta\phi = 0.1\times4.5^{\rm o}$ in the
forward region) and time resolution ($\sigma\simeq2.5$ ns).
In order to reduce the fake trigger and track probabilities as well as aging effects,
significant amount of shielding has been added. 
The subdetector is fully commissioned and giving an excellent performance.
\section{First Physics Results}
Over the course of 2001, significant progress was made in the development
and verification of algorithms to identify physics objects: electrons, muons,
jets, as well as to establish the electromagnetic and jet energy scales.
\subsection{QCD Physics}
\begin{figure}[h]
\begin{center}
  \begin{minipage}[t]{0.40\textwidth}
    \psfig{figure=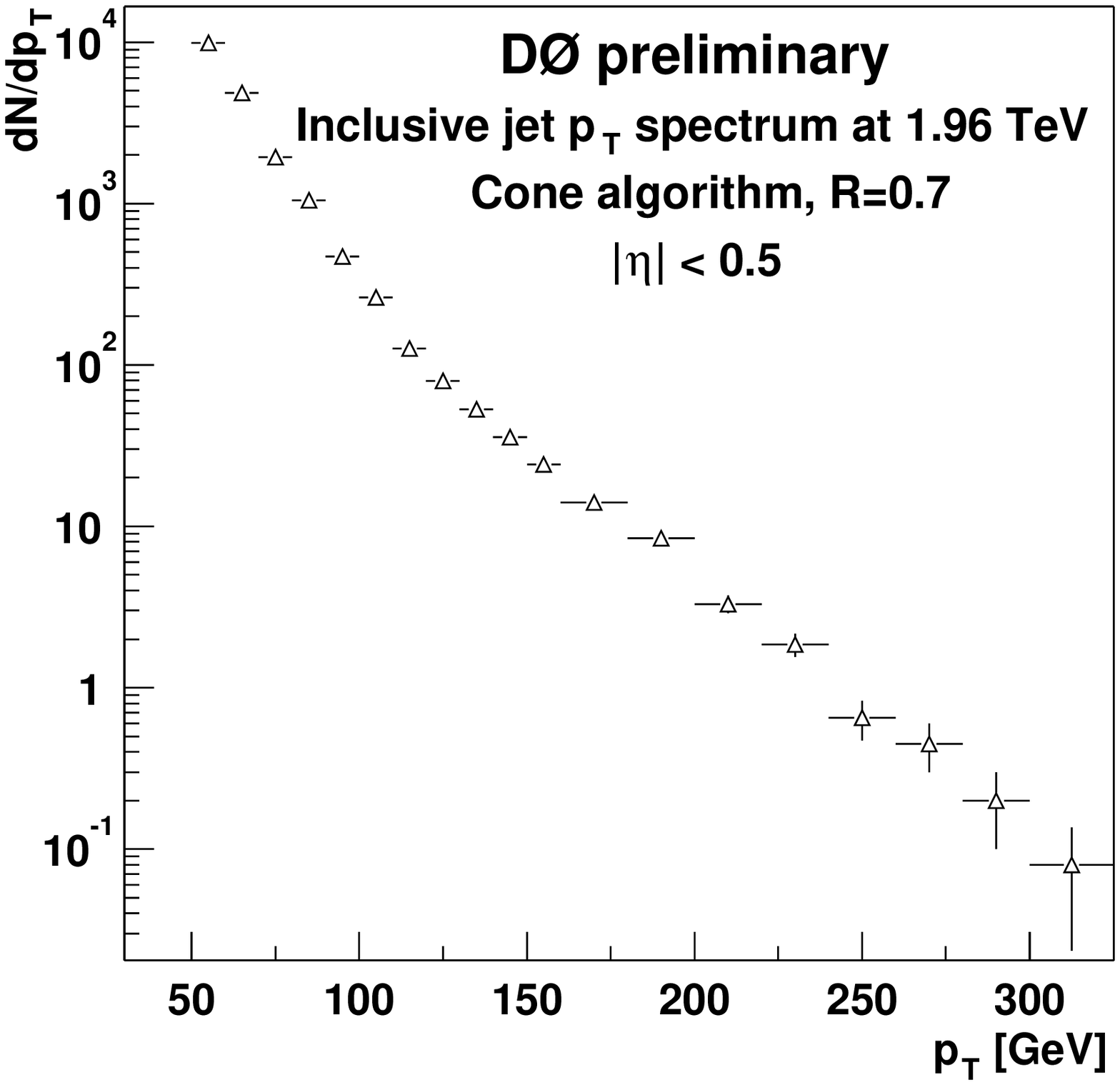,height=6cm}
    \caption{Inclusive jet p$_T$ spectrum. See text for details.} 
    \label{fig:qcdphysics}
  \end{minipage}
  \hspace{1cm}
  \begin{minipage}[t]{0.45\textwidth}
    \psfig{figure=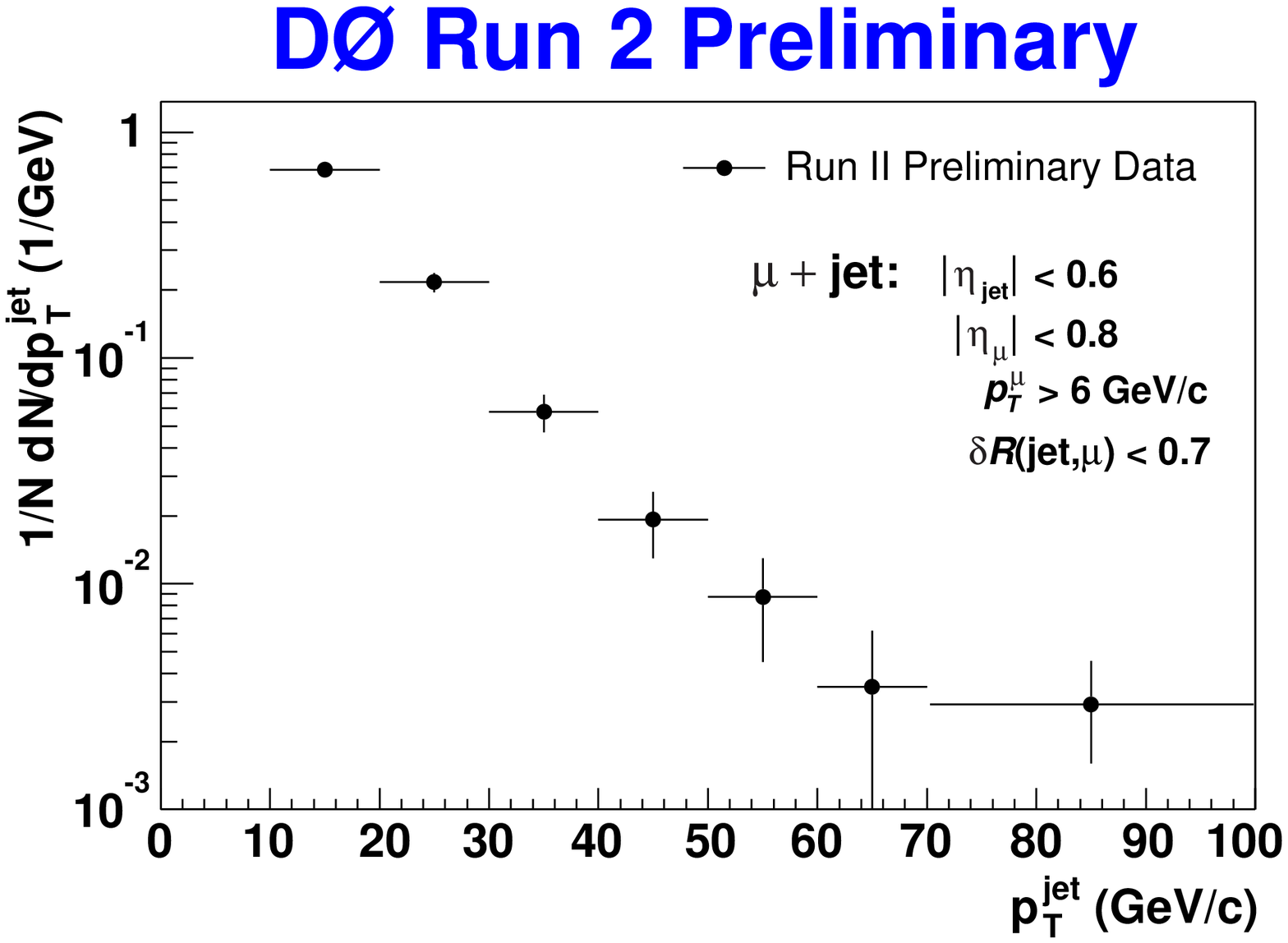,height=5.5cm,width=6cm}
    \caption{Inclusive jet p$_T$ spectrum for events with a muon within
             the jet cone. See text for details.}  
    \label{fig:bphysics}
  \end{minipage}
\end{center}
\end{figure}

The good performance of the calorimeter has allowed to perform
preliminary studies on jet physics, such as the inclusive jet
$p_T$ spectrum (Fig.~\ref{fig:qcdphysics}) or the dijet mass
spectrum. Jets are reconstructed in
the central region of the calorimeter ($\mid\eta\mid<0.5$) with a
cone algorithm with radius $R=0.7$ in $\eta-\phi$ space. These
spectra correspond to an integrated luminosity of $\sim1$
pb$^{-1}$ and are not fully corrected: there is a preliminary
correction for jet energy scale (but no unsmearing of resolution
effects) and there is no correction for trigger or selection
efficiencies.
\subsection{B-Physics}
The use of tracking allows to reconstruct particles into which
B hadrons typically decay: $K_{S}^{0}\rightarrow \pi^+\pi^-$,
$\Lambda^{0}\rightarrow p\pi^-$, $J/\Psi\rightarrow \mu^+\mu^-$, etc.
Another preliminary result is the measurement of the
inclusive jet $p_T$ spectrum (Fig.~\ref{fig:bphysics})
for events with a muon contained within a jet. 
This distribution corresponds to less than 0.2 pb$^{-1}$ and includes
trigger and reconstruction efficiencies, as well as jet energy
scale corrections. In order to obtain the inclusive b-jet $p_T$
spectrum, the b-jet content in the sample can be estimated by using the
large p$_T$ of the muon relative to the jet axis
to discriminate between direct $b\rightarrow\mu$ and
backgrounds ($c\rightarrow\mu$ and $\pi/K\rightarrow\mu$).
\subsection{Electroweak Physics}
\begin{figure}[h]
\begin{center}
  \begin{minipage}[t]{0.40\textwidth}
    \psfig{figure=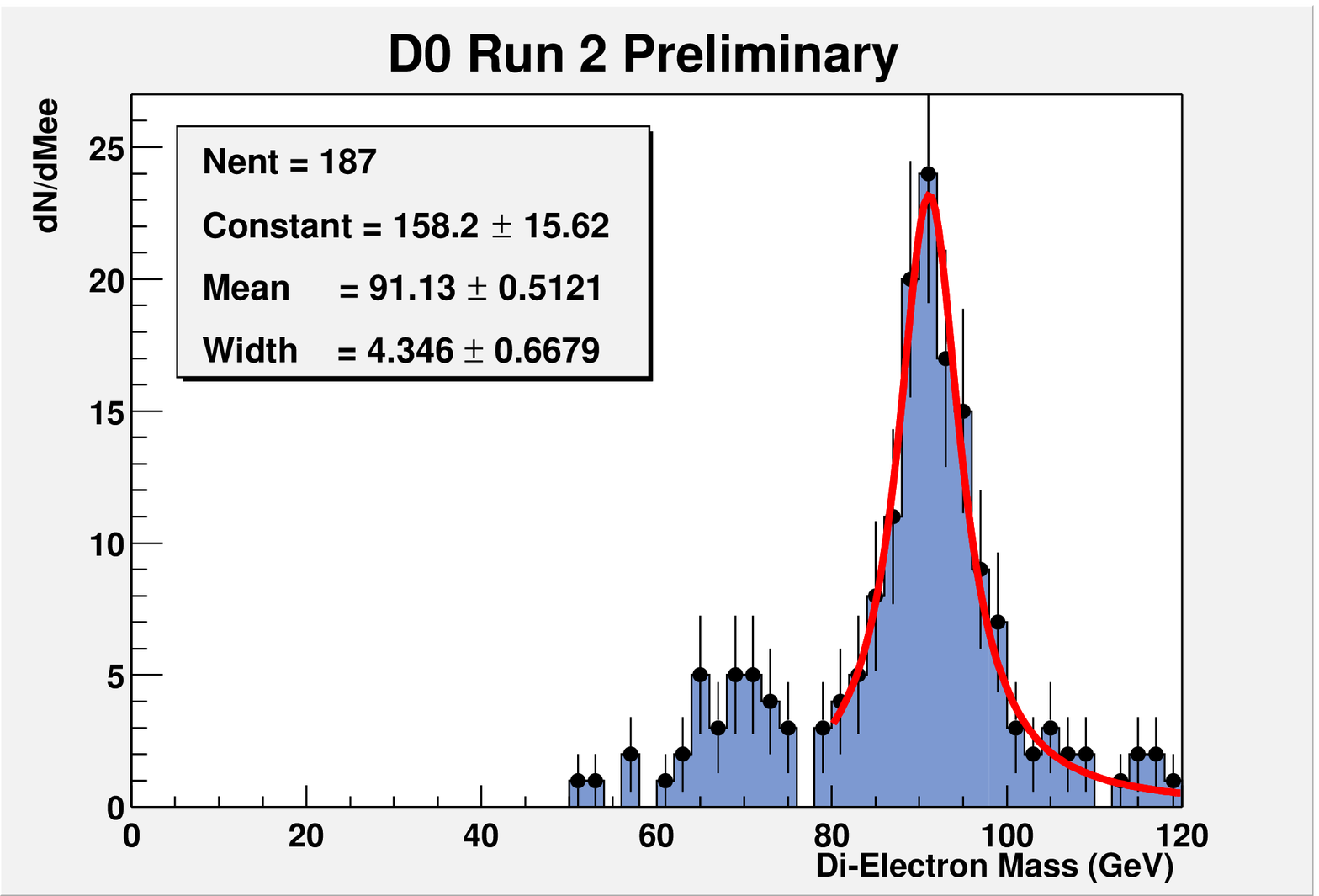,height=3.5cm,width=6cm}
    \caption{Di-electron invariant mass from $Z\rightarrow e^+e^-$
             candidates after energy scale corrections.} 
    \label{fig:ewphysics1}
  \end{minipage}
  \hspace{1cm}
  \begin{minipage}[t]{0.45\textwidth}
    \psfig{figure=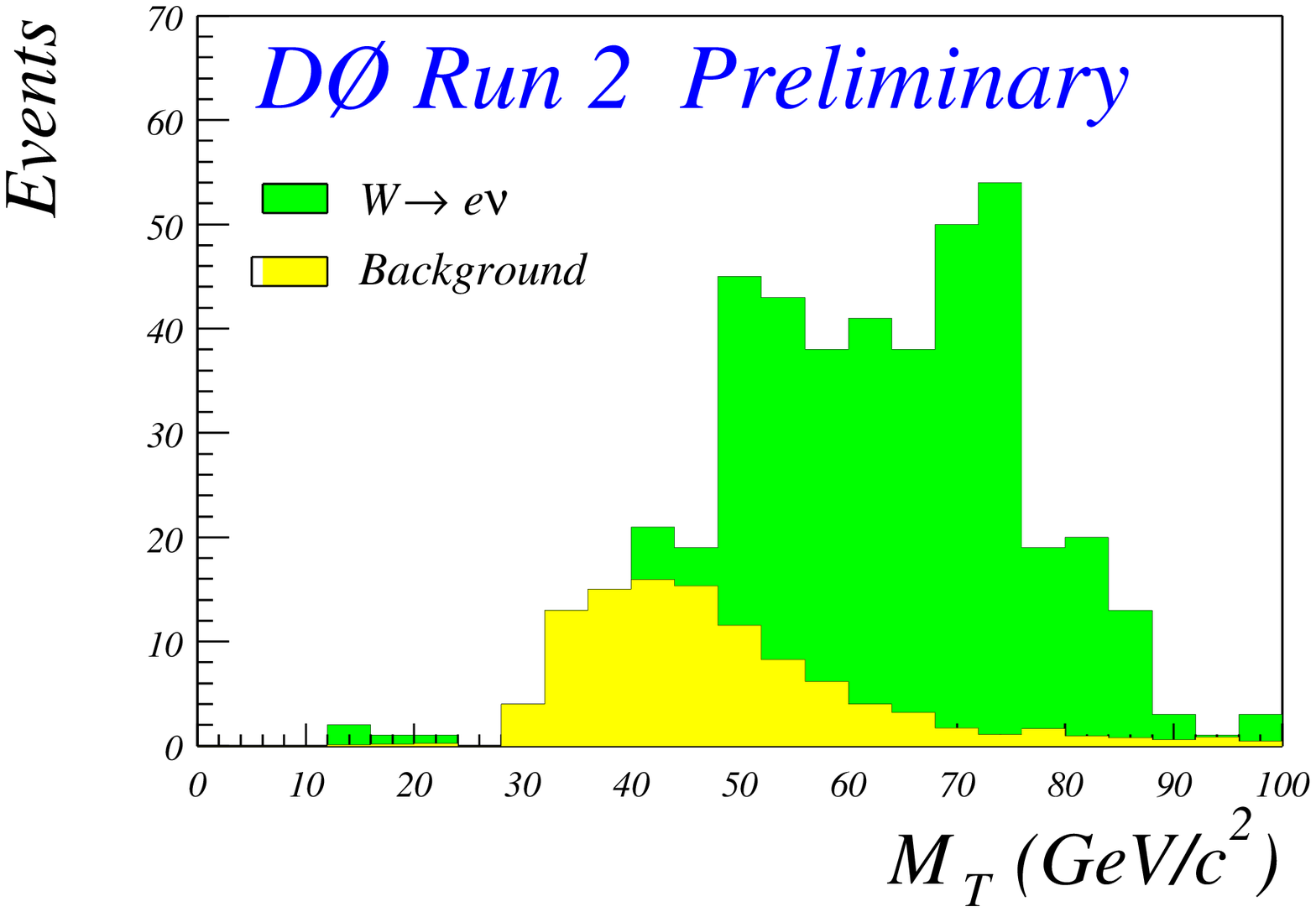,height=6.25cm,width=7cm}
    \caption{Transverse mass distribution for events
             containing a candidate $W\rightarrow e\nu$. The background, 
             estimated from the data, consists mainly of QCD events with fake electrons.}
    \label{fig:ewphysics2}
  \end{minipage}
\end{center}
\end{figure}

First preliminary results on electroweak physics include the
development of dedicated selections for Z and W bosons. The selection
of $Z\rightarrow e^+e^-$ events (Fig.~\ref{fig:ewphysics1}) requires two isolated
electromagnetic clusters in the calorimeter, with at least one of
them matched to a global track (using both the
SMT and the CFT) in order to reduce the dominant QCD
backgrounds involving photons and $\pi^0$s. This sample has been mainly
used to determine the absolute energy scale ($m_{ee}=m_Z$) as well
as to estimate the tracking efficiency. By requiring instead a
single electromagnetic cluster matched to a global track and large
missing $E_T$, a relatively clean sample of
$W\rightarrow e\nu$ candidates is selected, which shows in the transverse
mass distribution of the $e+\nu$ the
characteristic jacobian peak near $m_W$ (Fig~\ref{fig:ewphysics2}). 
Those candidate events with
additional jets ($W(\rightarrow e\nu)$+jets) will constitute the main background for
top-quark and Higgs boson analyses.
\subsection{New Phenomena Physics}
The D\O\ detector is particularly suited for searches for new phenomena
physics involving jets, isolated leptons and missing $E_T$.
Some preliminary results include the determination of the 
missing $E_T$ resolution from the inclusive di-electron sample
(an im\-por\-tant signature in searches for e.g. extra dimensions,
R-parity violating SUSY, etc), and searches for trileptons
and first generation leptoquarks, which have yielded some candidate events.
\section{Conclusions}
The Tevatron Run 2 started in March 2001, offering one of the most
exciting physics program of this decade. Over the course of 2001,
enormous progress has been made at D\O\ in terms of detector
commissioning. Preliminary performance results are encouraging and
indicate that the upgraded D\O\ detector will be able to fully
exploit the available physics opportunities. Despite the small
amount of integrated luminosity delivered so far, first physics
results are already emerging. The expected optimization of the
detector, trigger and data acquisition performances as well as
improvements in calibration, alignment, event selection and
reconstruction techniques give an impressive outlook.

\section*{Acknowledgments}
The author would like to thank Professor J. Tran Thanh Van and the conference
organizers for an stimulating and enjoyable conference.

\end{document}